\documentclass[structabstract]{aa}
\usepackage{graphicx}
\usepackage{longtable}
\usepackage{subfigure}
\usepackage{supertabular}
\usepackage{lscape}
\usepackage{hhline}
\usepackage{setspace}
\usepackage{threeparttable}
\usepackage{txfonts}
\usepackage{color}
\begin{document}
\title{FM\,047-02: a collisional pair of galaxies with a ring
\thanks{Based on observations made at the Gemini Observatory, under the identification number GS-2007A-DD-06.}}

\author{M. Fa\'{u}ndez-Abans\inst{1}
\and M. de Oliveira-Abans\inst{1,2}
\and A.C. Krabbe\inst{3}\\
P.C. da Rocha-Poppe \inst{4,5}
\and V.A. Fernandes-Martin\inst{4,5}
\and I.F. Fernandes\inst{4,5}
          }

\offprints{Max Fa\'{u}ndez-Abans; max@lna.br}

\institute{MCTI/Laborat\'{o}rio  Nacional  de  Astrof\'{\i}sica, Rua Estados Unidos, 154, Bairro das Na\c{c}\~{o}es,CEP 37.504-364, Itajub\'{a}, MG, Brazil \\
\email{mfaundez@lna.br, mabans@lna.br}
\and UNIFEI, Instituto de Engenharia de Produ\c{c}\~{a}o e Gest\~{a}o, Av. BPS 1303 Pinheirinho, 37500-903 Itajub\'{a}, MG, Brazil
\and Universidade do Vale do Para\'{i}ba - UNIVAP. Av. Shishima Hifumi, 2911 - Urbanova
    CEP: 12244-000 - S\~{a}o Jos\'{e} dos Campos, SP, Brazil.
             \email{angela.krabbe@gmail.com}
\and  UEFS, Departamento de F\'{i}sica, Av. Transnordestina, S/N, Novo Horizonte, Feira de Santana, BA, Brazil, CEP 44036-900
\and UEFS, Observat\'{o}rio Astron\^{o}mico Antares, Rua da Barra, 925, Jardim Cruzeiro, Feira de Santana, BA, Brazil, CEP 44024-432 
\\
\email{paulopoppe@gmail.com, vmartin1963@gmail.com, irafbear@gmail.com}}  

   \date{Received 22 Juanary 2013 / Accepted 17 July 2013}

 
  \abstract
{}
{We investigate the nature of the galaxy pair \object{FM\,047-02}, which has been proposed as an archetype of the Solitaire types of peculiar (collisional) ring galaxies.}
{The study is based on long-slit spectrophotometric data in the range of 3500-9500 $\AA$ obtained with the Gemini Multi-Object Spectrograph at Gemini South (GMOS-S). The absorption spectra were used to determine the radial velocity. The stellar population synthesis code STARLIGHT was used to investigate the star formation history of the galaxy pair FM\,047-02.}
{During the whole long-slit signal, the spectra of both galaxies resemble those of an early-type galaxy. Both objects are dominated by an old stellar population of $2\times10^{9} <\rm t \leq 13\times10^{9}$ yr and a small, but a non-negligible fraction (about 13\%) of young stars of $\rm t \leq 5\times10^{7}$ yr
are estimated to contribute to the optical flux for NED02.  Both observed small radial-velocity differences, and the structures around NED01 suggest an ongoing tidal interaction of both galaxies.}
{The spectroscopic results and the morphological peculiarities of NED01 can be adequately interpreted as continuing off-center stage of interaction with the companion galaxy NED02. The core of NED01 and NED02 are in counter-rotating motion. The visual appearance of NED01, its kinematical properties, the smooth distribution of material in the ring, and its off-centered nucleus, characterize it as a true archetype of a Solitaire-type ring galaxy in an advanced stage of ring formation.}
{}


\maketitle
\titlerunning{FM\,047-02}
\authorrunning{Fa\'undez-Abans et al.}

%

\section{Introduction}

Today it is well accepted that most galaxies experience several collisions and/or tidal interactions over the course 
of their lifetimes. Some of these interactions may be strong enough to profoundly alter their structure and accelerate their evolutionary process. 
Thus, collisions and interactions are now generally believed to be one of the primary drivers of galaxy evolution and morphological transformation of the structure of a galaxy. The processes of 
galaxy formation and evolution are intimately connected to star formation, as are a variety of other 
characteristics related to galaxies, such as the creation of heavy elements in the Universe by stellar
evolution and the formation of galactic structures, planetary systems, production and distribution of cosmic rays, 
among others. One of the possible results of collisions and tidal interactions in galaxies are rings triggered 
in galactic disks.

It is noteworthy here to clarify that not all ring galaxies have the same origin. They can be divided into two main groups: the normal Ring galaxies (NRG) and the peculiar Ring galaxies (pRG) (Fa\'{u}ndez-Abans \& de Oliveira-Abans \cite{foa98} hereafter FAOA). The rings of NRG result from resonances between the gravitational field of structures,  such as bars, ovals, or spiral arms, and their oscillation about the circular orbits in a galactic disk. The pRGs are the result of tidal interaction between two galaxies, a collisional scenario that ends in the merger and transmutation of the objects (Appleton \& Struck-Marcell \cite{apple1996}). The pRGs show a wide variety of ring and bulge morphologies and are classified by  Fa\'{u}ndez-Abans \& de Oliveira-Abans (\cite{foa98}) into five families, following the general behavior of galaxy-ring structures. Eight morphological subdivisions are highlighted for these categories.  One of them is a basic structure called ``Solitaire". 
The Solitaire pRG is described there as an object with the bulge on the ring or very close to it, resembling a one-diamond finger ring (single knotted ring). In these objects, the ring generally 
looks smooth and almost thin on the opposite side of the bulge (for archetypes see \object{FM\,188-15/NED02}, \object{AM\,0436-472/NED01)}, \object{ESO 202-IG45/NED01}, and \mbox{\object{ESO 303-IG11/NED01})}.
There are still a few statistical works on this type of object, and details of the collisional process behind their formation is still unknown. 

In this work, we report a first study with spectroscopic observations of the tidally disturbed galaxy NED01, a member of the galaxy pair \object{FM\,047-02} (\object{ESO 046-IG10}; \mbox{\object{AM\,2021-724}};
 see also Arp \& Madore \cite{am86}) and its companion NED02, based on long-slit spectrophotometric data obtained with the Gemini GMOS Spectrograph at the Gemini Observatory \mbox{(ID: GS-2007A-DD-06)} in Chile. 
A value of $H_{\rm o}$ = 70 km s$^{-1}{\rm Mpc}^{-1}$,\, $\Omega_{matter} = 0.27$ and\, $\Omega_{vaccum} = 0.73$ have been adopted throughout this work (Freedman et al. \cite{f2001}; Astier et al. \cite{a2006}, see also Spergel et al. \cite{s2003}). Table~\ref{table1} presents this work's new results for both galaxies.

\section{Observations and data reduction}

\begin{figure}
\centering
\resizebox{80 mm}{!}{\includegraphics[clip]{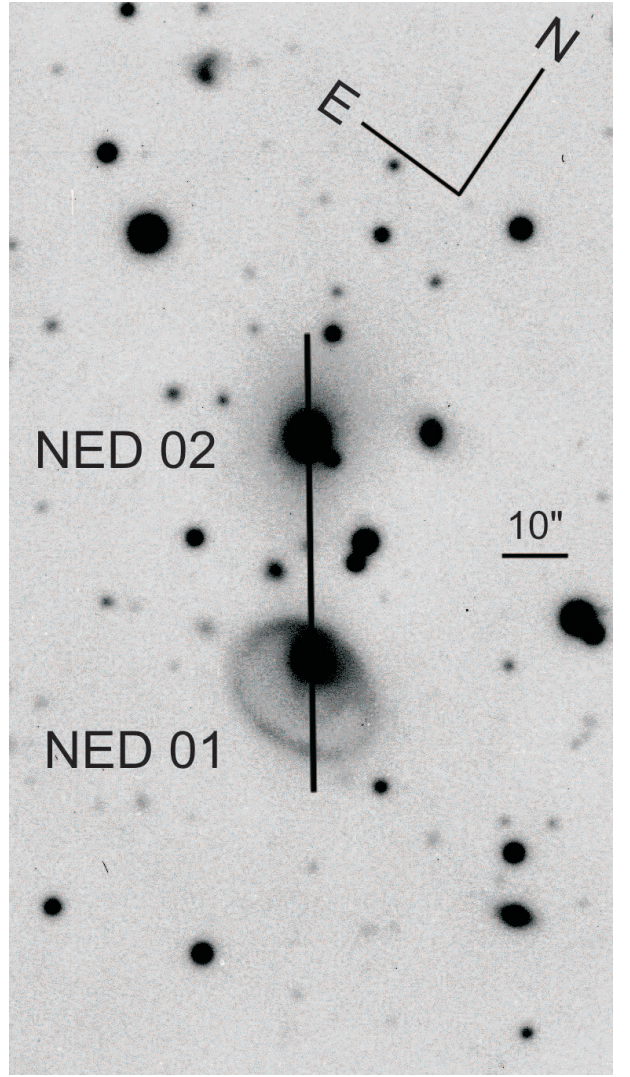}}
\caption{The Solitaire galaxy NED01 and its companion NED02. 
Original optical GMOS pointing image in the r-G0303 filter, with contrast to highlight the main morphological features of the galaxies. The PA = 28$^{\circ}$ slit position  is displayed.}
\label{fig_01}
\end{figure}

The FM\,047-02 galaxy pair is composed of a ringed galaxy, NED01 (2MASX J20262950-7236443); and an elliptical-like object, NED02 (2MASX J20263396-7236083). 
Fig.~\ref{fig_01} displays both galaxies from the telescope pointing 5 minute-exposure GMOS image (r-G0303 filter with an effective wavelength of 630 nm). 
The spectroscopic observations were performed with the 8.1-m Gemini South telescope, Chile 
\mbox{(ID program GS-2007A-DD-06)}. We used the GMOS spectrograph in long-slit mode 
(Hook et al. \cite{h04})\footnote{A description of the instrument can be found at 
http://www.gemini.edu/sciops/instruments/gmos}. The long slit used has an entrance on the plane of the sky of 
1.5 \arcsec $\times$ 375 \arcsec. A grating of 400 lines/mm (R400+G5325, with GG455-G0329 blocking filter) centered on 676.5 nm was used. The data were binned by two pixels in both the
 spatial and spectral dimensions, 
producing a spectral resolution of $\sim$5.1\,$\AA$ $FWHM$ sampled at \mbox{0.68 $\AA$ pix$^{-1}$}. 
The seeing throughout the observations was 1\farcs5. The binned pixel scale was 0\farcs145 pix$^{-1}$. 
The wavelength range is $\sim$3\,500-9\,500 $\AA$. The star LTT\,7379 was used for extinction and flux calibrations. It is a tertiary standard from Baldwin \& Stone (\cite{BS}), as revised by Hamuy et al. (\cite{H92}, see also Hamuy et al. \cite{H94}). The long-slit spectra were taken at one position angle on the sky, PA = 28$\degr$, to cover both objects in one shot and across the center of both objects.

The standard Gemini-$IRAF$ routines were used to carry out bias subtraction, flat-fielding, and cosmic ray hits. The data were then calibrated in wavelength with an accuracy of \mbox{$\leqslant$ 0.3 $\AA$}.
The 2-D spectra was then extracted into 1-D spectra which were sky-subtracted and binned in the spatial dimension. The binned 1-D spectra were flux-calibrated using the spectrophotometric standard star LTT\,7379. 

\begin{table}
\caption{Data on FM\,047-02 galaxies.}
\label{table1}
\centering
\begin{tabular}{llll}
\hline \hline
Parameter &  NED01 & NED02 & Ref.\\
\hline
 R.A. (2000)  &  20 26 29.5 & 20 26 33.9 & NED \\
Dec. (2000) & \hspace{-0.1cm}-72 36 44.5  &  \hspace{-0.1cm}-72 36 08.2  & NED \\
Morphology & Solitaire & E4: & this work \\
pRG-family  & Solitaire-like &  & FAOA   \\
$z$ & 0.0767  & 0.0759 & this work \\
$V$(km\,s$^{-1}$)$^{({\it *})}$ & 22\,994 $\pm$25 & 22\,754 $\pm$20& this work \\
Distance (Mpc) & 322  & 322  & this work \\
$\sigma_{v}$ (km/s) & 387 & 285 & this work \\
Mass (lower limit) & 3.97$\times10^{11}$M$_{\sun}$ & 1.93$\times10^{11}$ M$_{\sun}$& this work \\
R$_{eff}$ (adopted) & 12\farcs7 & 11\farcs5 & this work \\
\hline
\end{tabular}
\begin{tablenotes}
\item[a]{{(\it *)}: non relativistic velocity using the standard formula (Lang \cite{lang1999}).}
\end{tablenotes}
\end{table}


\section{Kinematics}

\begin{figure}[h]
\begin{center}$
\begin{array}{cc}
\resizebox{\hsize}{!}{\includegraphics[]{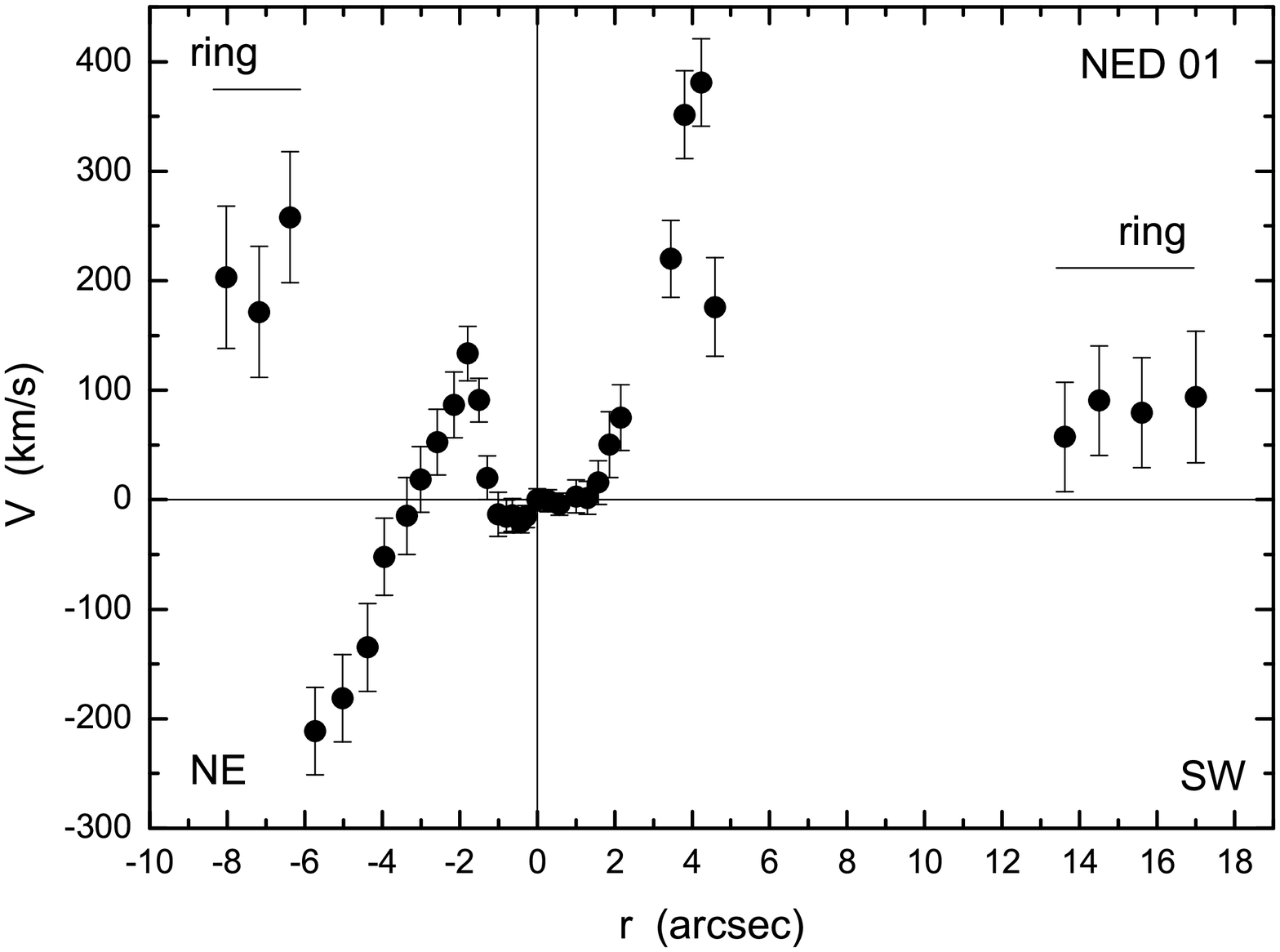}}\\
\resizebox{\hsize}{!}{\includegraphics[]{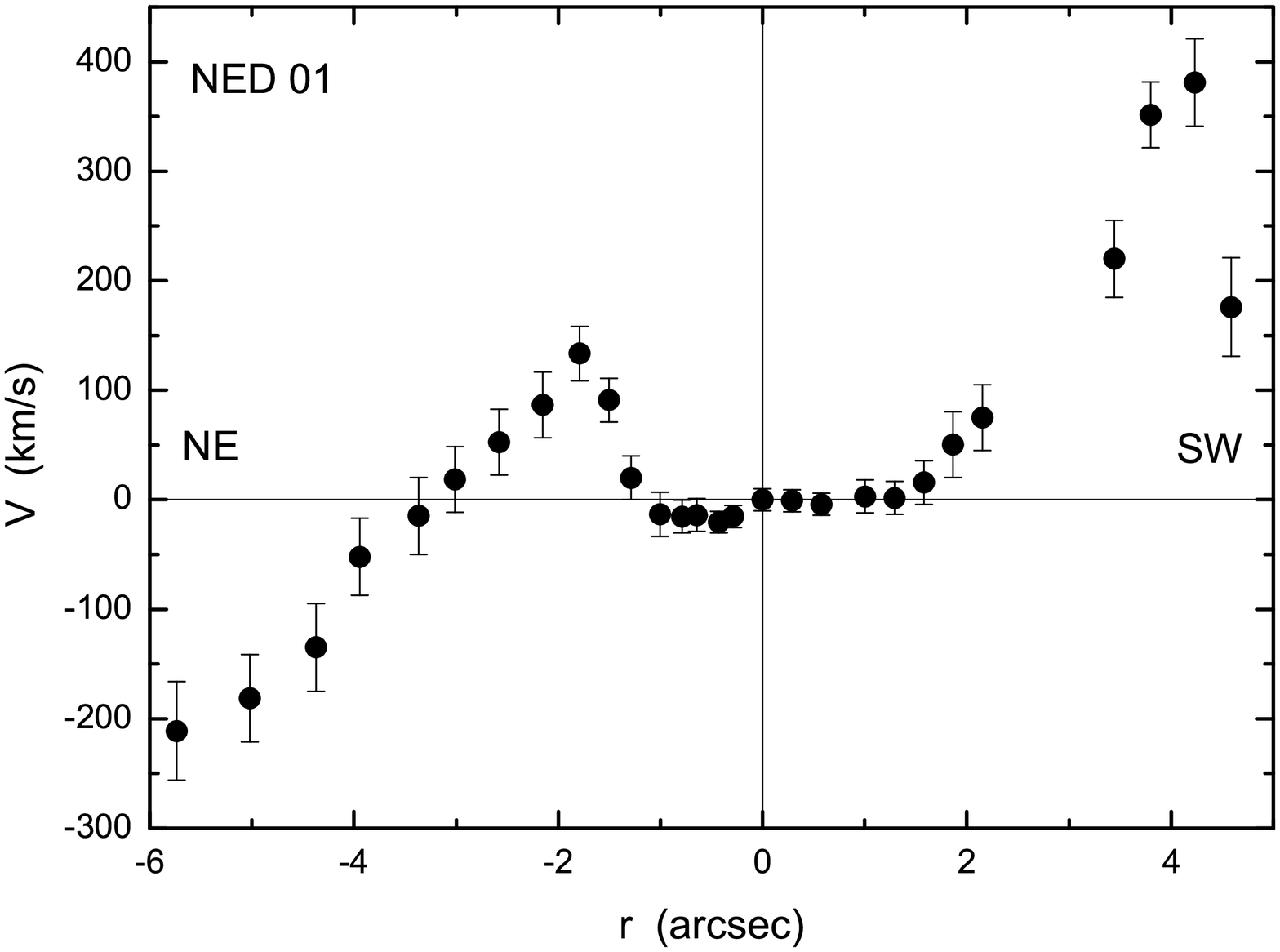}}
\end{array}$
\end{center}
\caption{From top to bottom:  (a) the NED01 rotation profile of the nucleus, bulge and ring sections, the total long-slit $-$10\arcsec to $+$19\arcsec distribution (which shows a U-shaped rotation profile at the center), (b) the central $-$6\arcsec to $+$5\arcsec nuclear and bulge features.}
\label{fig_02}
\end{figure}

The radial velocity for each galaxy was estimated by the cross-correlation technique. Using $IRAF/RVXCOR$ we cross-correlated our observed spectra with three galaxy and star templates with high signal-to-noise. These results were checked with the composite absorption-line template ``fabtemp97" distributed by the RVSAO\footnote{The RVSAO IRAF (Radial Velocity Package for IRAF) 
external package was developed at the Smithsonian Astrophysical Observatory. Full documentation of this software, including numerous examples of its use, is online 
at http://tdc-www.harvard.edu/iraf/rvsao/.}/IRAF external package. We adopted the redshift value from the best high correlated coefficient template. Figures~\ref{fig_02}-\ref{fig_03} display the distribution of the radial velocity for NED01 and NED02, respectively. Figure~\ref{sintese1e2} presents a sample of integrated spectra in rest frame and the model stellar population espectrum in the range of 4\,600-6\,500 $\AA$ 
for NED01 and NED02, respectively (see Section \ref{sintese}  for more details).
The spectra had significant fringing in the red (7\,000-9\,500 $\AA$), and no emission 
lines in both spectra were detected along the entire slit, even in the selected regions of the NED01 ring. This region  of the spectra with  significant fringing was not used in our analysis.


\subsection{NED01}

The spectral profile resembles that of an early-type object. The velocity found is V= 22\,994 km\,s$^{-1}$ ($z$=0.0767).  Fig.~\ref{fig_02} displays  NED01's  distribution of radial velocities along the whole object (upper panel) and the central region (lower panels). The errors of the individual velocity measurements 
do not exceed 10 km\,s$^{-1}$ in the central region and increase to 20$-$40 km\,s$^{-1}$ on its periphery.  The two off-centered main ring sections ($-$6\arcsec to $-$8.5\arcsec and 13\arcsec to 17.5\arcsec, respectively) and the loop-ring (close to $+$4\arcsec north section) are receding from us.

Clearly NED01 is a tidally perturbed object. A U-shaped profile is displayed in the $\pm$4\arcsec distribution section of the radial velocities. This shape appears on interacting binary-disturbed elliptical galaxies (see Borne \cite{b1990}; Borne \& Hoessel \cite{bh1985},\cite{bh1988}; Bender et al. \cite{bpn1991}; Madejsky \cite{mad1991} 
and Madejsky et al. \cite{madall1991}). The physical interpretation  of Borne et al. (\cite{borne1994}) 
is that there is a tidal coupling between the orbit of the companion and the resonant prograde rotating stars in the kinematically disturbed galaxy (Borne \cite{bor1988}; Borne \& Hoessel \cite{bh1988}; Bacells et al. \cite{bacells1989}). This U-shape profile in NED01 is thus direct observational evidence of tidal coupling, hence a direct observational signature of tidal friction in action in this pair of galaxies.

Inspecting the upper panel of Fig.~\ref{fig_02}, we realize that there are several kinematical subsystems throughout the regions comprised by the slit: (1) the U-shaped base of the rotation profile; (2) the wings of the U-shaped rotation profile; (3) a rigid rotator (from $-$6\arcsec to $-$2\arcsec \,NE-section; although internally wavy, the one on the rotator's left is approaching, while the one on its right is receding from us); and (4) the ring sections. The lower panel of Fig.~\ref{fig_02} shows the observed radial velocity from $-$4.5\arcsec to 4.5\arcsec, and the enlarged U-shape.

\subsection{NED02}

The spectrum of NED02 shows characteristics of an early-type galaxy, which are typical of elliptical galaxies, like the dwarf ellipticals that populate the Virgo cluster of galaxies (van Zee et al. \cite{zee2004}). 
The velocity that we found is $V$= 22\,754 km\,s$^{-1}$ ($z$ = 0.0759), which is in good agreement with the previously reported value $z$ = 0.0756 on NED.

\begin{figure}
\centering
\resizebox{\hsize}{!}{\includegraphics[clip]{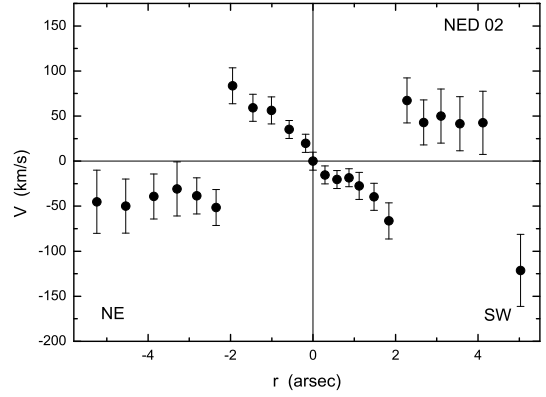}}
\caption{The NED02 observed distribution of the rotation profile of the nuclear, bulge, and external features, respectively.}
\label{fig_03}
\end{figure}

Figure~\ref{fig_03} displays the distribution of radial velocities of NED02 measured along all the object. The errors of the individual velocity measurements do not exceed 10 km s$^{-1}$ in 
the central region and increase to 20$-$30 km s$^{-1}$ on its periphery. In this figure, the rotation curve comprising $\pm$\,2\arcsec around the nucleus is similar to those of galaxies with kinematically decoupled regions.
(Balcells \& Gonz\'{a}lez \cite{bg98}, see also Hau \& Thomson \cite{ht94} and De Rijcke et al. \cite{rdzh04}). Inspecting  Fig.~\ref{fig_03}, we realize that there are two main kinematical 
subsystems along the observed slit: (1) a core-bulge rigid rotator at $\pm$\,2\arcsec, with almost the same angular velocity as suggested by the average inclination of their linear profile (its SW part is approaching while its NE part is receding from us); and (2) the main body with a flat rotation curve around $\pm$ 50 km\,s$^{-1}$ (r $>$ $\pm$\,2\arcsec) decoupled from the core-bulge section. The NE-section is approaching while the SW-section is receding from us. The latter is the ``contact region" between NED02 and NED01. Adopting the $b/a = 0.84$ ratio from the pointing image, the morphology of NED02 suggests that it is an  elliptical galaxy, maybe an E4.

\section{Stellar population synthesis}
\label{sintese}

\begin{figure*}
\subfigure{\includegraphics*[angle=-90,width=\columnwidth]{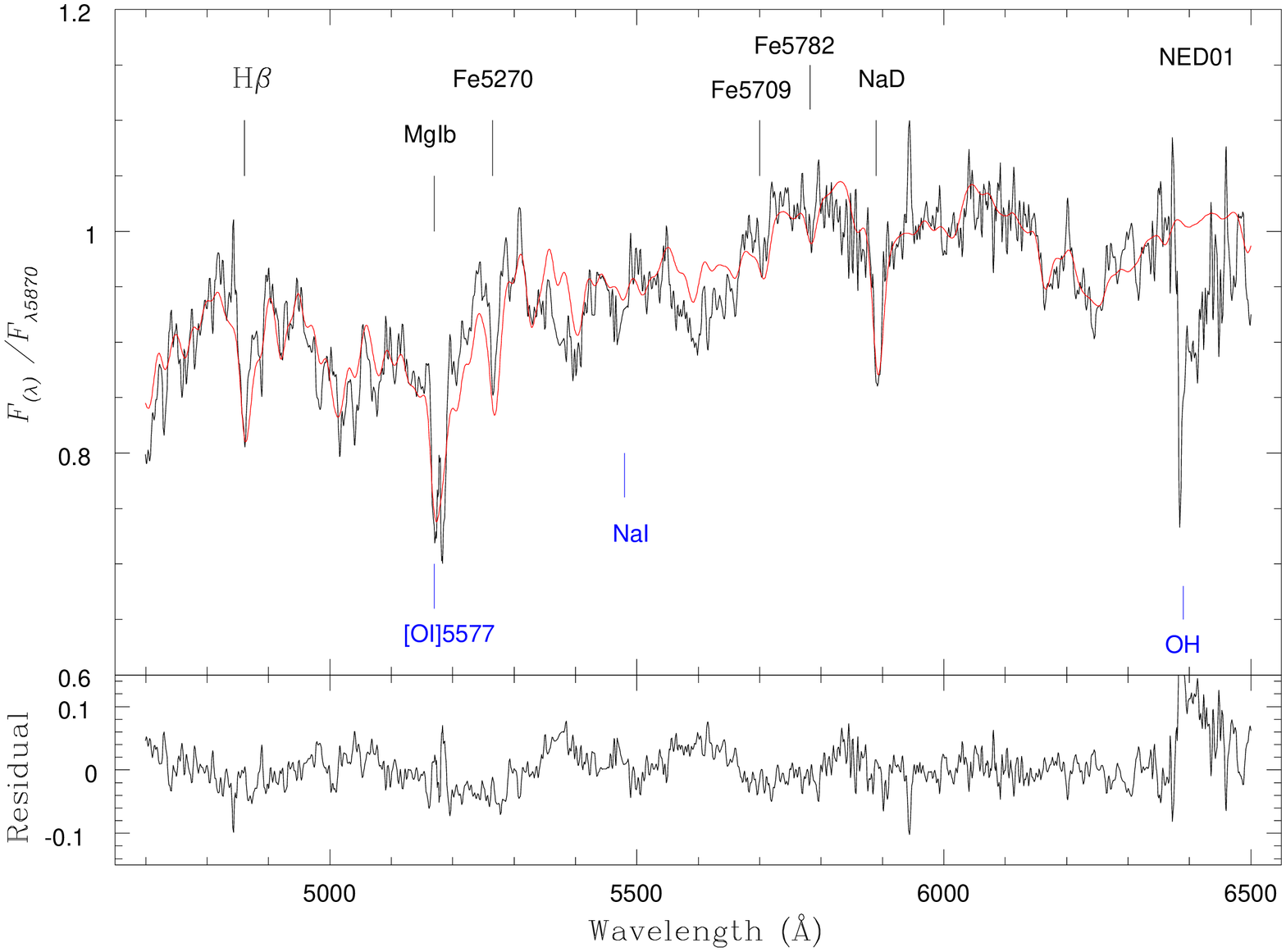}}
\subfigure{\includegraphics*[angle=-90,width=\columnwidth]{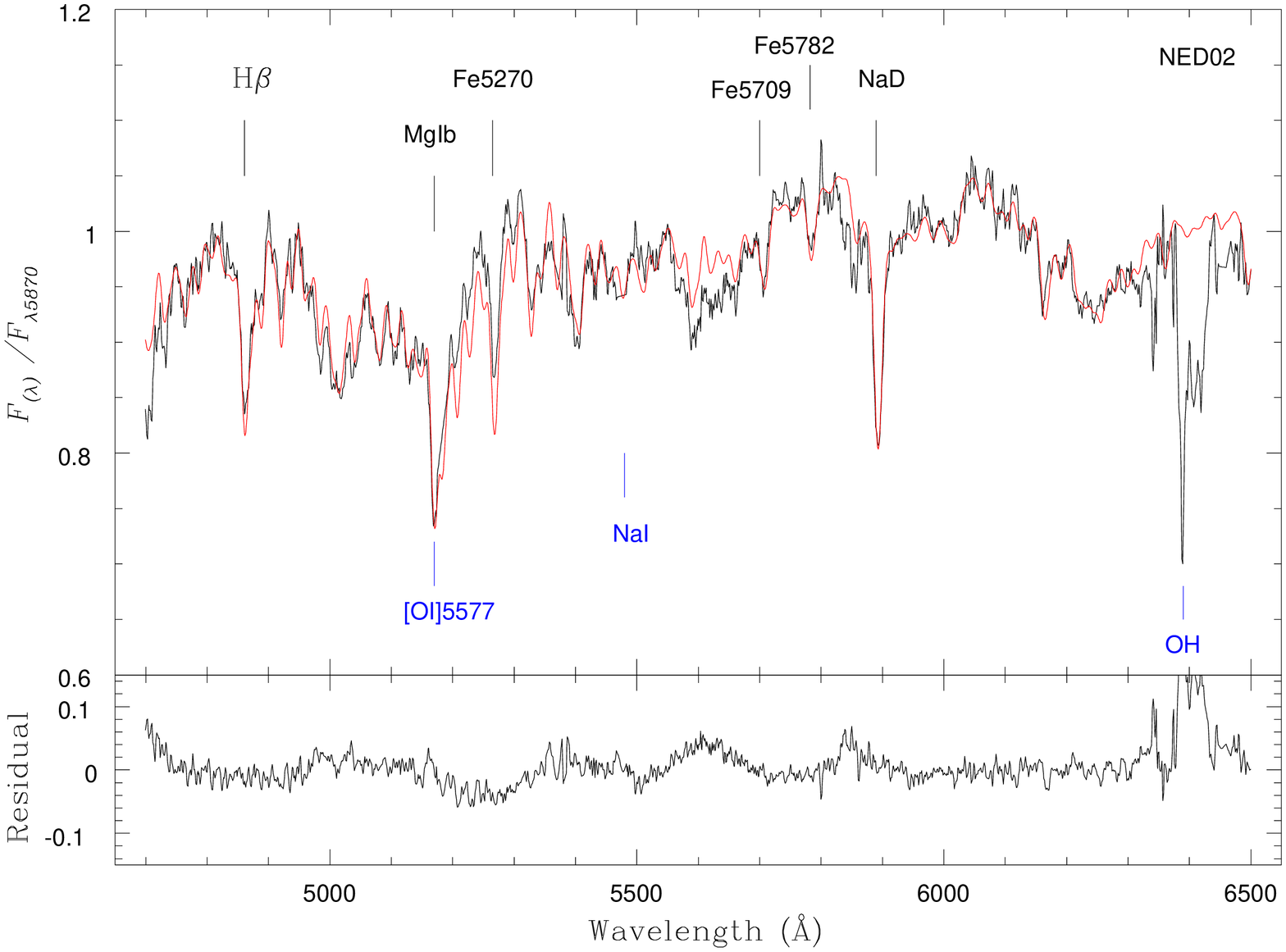}}
\caption{Stellar population synthesis for NED01 and NED02 . Observed spectra corrected for reddening (black line) and the synthesized spectrum (red line). The main absorption features have been identified and the most significant telluric lines and bands marked. The spectra are at rest frame.}
\label{sintese1e2}
\end{figure*}

Detailed study of the star formation in tidally perturbed galaxies provides important 
information, not only on the age distribution along their stellar population components, but also on 
several aspects related to the interacting process and to its effects on the properties 
of the individual galaxies and their evolution.

To investigate the star formation history of NED01 and NED02, we used the stellar population synthesis code {\sc STARLIGHT} (Cid Fernandes et al.\cite{cid04,cid05}; Asari et al.\cite{asari07}).
This code is extensively discussed in Cid Fernandes et al. (\cite{cid04,cid05}), and is built upon computational techniques originally developed for empirical population synthesis with additional 
ingredients from evolutionary synthesis models. This method was also used by Krabbe et al. (\cite{krabbe2011}) and has been successful in describing the stellar population in interacting galaxies.

The code fits an observed spectrum $O_{\lambda}$ with a combination of $N_{\star}$ single stellar populations (SSPs) from the Bruzual \& Charlot (\cite{bruzual03}) models. 
These models are based on a high-resolution library of observed stellar spectra, which allows for detailed spectral evolution of the SSPs at a resolution of 3\,\AA\, across 
the wavelength range of 3\,200-9\,500 $\AA$ with a wide range of metallicities.  We used the Padova 1994 tracks as recommended by Bruzual \& Charlot (\cite{bruzual03}), 
with the initial mass function of Salpeter (\cite{salpeter55}) between 0.1 and 100 $M_{\sun}$. Extinction is modeled by {\sc STARLIGHT} as due to foreground dust, using the Large Magellanic Cloud average reddening law of Gordon et al. (\cite{gordon03}) with  R$_V$= 3.1, and parametrized by the V-band extinction  A$_V$. The SSPs used in this work cover 15 ages, t = 0.001\,, 0.003\,, 0.005\,, 0.01\,, 0.025\,, 0.04\,, 0.1\,,  0.3\,, 0.6\,, 0.9\,, 1.4\,, 2.5\,, 5\,, 11\,, and 13 Gyr, as well as three metallicities, Z = 0.2 Z$_{\sun}$, 1 Z$_{\sun}$, and 2.5 Z$_{\sun}$, added to the $N_{\star}$ = 45 SSP components. 

The fitting was carried out using  a simulated annealing plus Metropolis scheme, which searches for the minimum  \mbox{$ \chi^{2} = \sum_{\lambda}[(O_{\lambda} - M_{\lambda})w_{\lambda}]^{2}$} where $w_{\lambda}$ is the error in $O_{\lambda}$ and $M_{\lambda}$ the model spectrum. The model spectrum M$_{\sun}$ is obtained by the equation

\begin{equation}
M_{\lambda} = M_{\lambda\,0}\biggl[\sum^{N_{\star}}_{j=1} x_{j} b_{j},_{\lambda} r_{\lambda}\biggr] \otimes G(v_{\star}, 
\sigma_{\star}) 
\end{equation}
where $b_{j},_{\lambda} r_{\lambda}$ is the reddened spectrum of the ${j}$th SSP normalized at
$\lambda _{0}$; $r_{\lambda} = 10^{0.4(A_{\lambda}- A_{\lambda_0})}$ is the reddening term; M$_{\lambda 0}$ is the synthetic
flux at the normalization wavelength; $\vec{x}$ is the population vector;
$\otimes$ denotes the convolution operator; and G($v_{\star}, \sigma_{\star}$) is the Gaussian
distribution used to model the line-of-sight stellar motions. It is centred at velocity $v_{\star}$ with dispersion $\sigma_{\star}$. Bad pixels and spurious features are masked out by fixing $w_{\lambda}=0$, such as the region around Mg$_{1}$ that was affected by sky residuals. We keep the kinematical parameters fixed during the fit, assuming the values of the velocity dispersion  obtained  from the cross-correlation.

Prior to the modeling, the SSPs models were  convolved to the same resolution of the observed spectra:
the observed spectra were shifted to its rest frame, corrected for foreground Galactic reddening of $E(B-V)=0.053$ mag taken from Schlegel et al. (\cite{sc98}), and normalized to $\lambda\,5870\,$\AA. The observed spectra were resampled in steps of $\Delta\, \lambda=1\, \AA$, the same as for the SSPs models. 
We synthesized the stellar population for the integrated spectrum (all galaxy) of NED01 and NED02.

Figure~\ref{sintese1e2} shows the observed spectra corrected for reddening, the  model stellar population spectra, and the residual spectra  for NED01 and NED02. The results of the synthesis are summarized in Table \ref{synt_table} for the  integrated spectrum of  each galaxy, stated as the perceptual contribution of each base element to the flux  at $\lambda$\,5\,870\, \AA. Following the prescription of Cid Fernandes et al. (\cite{cid05}), 
we defined a condensed population vector by binning the stellar populations according to the flux contributions into young, \mbox{$x_{\rm Y}$ ($\rm t \leq 5\times10^{7}$ yr)};
intermediate-age,  \mbox{$x_{\rm I}$ ($ 5\times10^{7} <\rm t \leq 2\times10^{9}$ yr)}; and 
old, \mbox{$x_{\rm O}$ ( $2\times10^{9} <\rm t \leq 13\times10^{9}$ yr)} components. The same bins were used to
represent the mass components of the population vector ($m_{\rm Y}$, $m_{\rm I}$, and
$m_{\rm O}$). The metallicity (Z), one important parameter for characterizing the stellar population content, is weighted by the light fraction. The quality of the fitting result is measured by the parameters 
$\chi^{2}$ and  $adev$. The latter gives the perceptual mean deviation $|O_{\lambda} - M_{\lambda}|/O_{\lambda}$ over all fitted pixels, where $O_{\lambda}$ and $M_{\lambda}$ are the observed and model spectra, respectively.

The results indicate that NED01 is completely dominated by an old stellar population of $2\times10^{9} <\rm t \leq 13\times10^{9}$ yr. For NED02, a small fraction of about 13\% of young stars  with ages $\leq 5\times10^{7}$ yr contribute to the optical flux at $\lambda\,5870\,$\AA.  As can be seen in  Fig.~\ref{sintese1e2}, the  synthetized spectra show small disagreement (on average $<  5\%$) with observed spectra in both galaxies, such as at about  $\lambda$\,5\,600 \AA. Besides this, no emission lines were detected in the spectra of NED02, which weakens the conclusion about the  small contribution of young star population from the synthesis for NED02. Concerning about it, we used only SSPs with old ages and different metallicities in the synthesis fitting.  These results are shown  in Table \ref{synt_table}. The residuals found in this case almost have the same magnitude as the one found when SSPs with different ages are used. However, it resulted in very low extinction values of $A_{v}$.  The error of $A_{v}$ estimations is on the order of 0.1 mag  (Cid Fernandes et al. \cite{cid2013}).  According to Cid Fernandes et al. (\cite{cid05}), in the fitting  the extinction $A_{v}$ is not constrained to be positive. Many reasons are pointed out by the authors for this choice, among them, that some objects may indeed require bluer SSP spectra than  those in the base and that  the observed light  may contain a scattered component, which would induce a bluening of the spectra not taken into account by adopted pure exctinction law (for more details see in quoted article). Similar results were found for HRG\,2304, a Solitaire galaxy candidate in an early stage of ring formation, and its companion \object{AM 1646-795} (Wenderoth et al. \cite{wend11}).

\begin{table*}
\caption{Stellar-population synthesis results (see text).}
\label{synt_table}
\begin{tabular}{lrrrrrrrrrr}
 \\
\noalign{\smallskip}
\hline
\hline
\noalign{\smallskip}
Galaxy &  \multicolumn{1}{c}{$ x_{\rm Y}$} &  \multicolumn{1}{c}{$x_{\rm I}$} &  \multicolumn{1}{c}{$x_{\rm O}$}
 & \multicolumn{1}{c}{$ m_{\rm Y}$}& \multicolumn{1}{c}{$ m_{\rm I}$} &
 \multicolumn{1}{c}{$m_{\rm O}$}& 
  \multicolumn{1}{c}{$Z_{\star}$[1]} &
 \multicolumn{1}{c}{$ \chi^{2}$} & 
 \multicolumn{1}{c}{$\rm adev$} & \multicolumn{1}{c}{$\rm A_{v}$}
 \\
 & \multicolumn{1}{c}{(per cent)}  &
 \multicolumn{1}{c}{(per cent)} &  \multicolumn{1}{c}{(per cent)}
 & \multicolumn{1}{c}{(per cent)} & 
 \multicolumn{1}{c}{(per cent)}
   &
 \multicolumn{1}{c}{(per cent)}& 
  \multicolumn{1}{c}{} &
 \multicolumn{1}{c}{} & 
 \multicolumn{1}{c}{(mag)}
 \\
\hline
\noalign{\smallskip}
NED01	  &   0.0     &   0.0  &   100.0  &   0.0  &   0.0  &   100.0  & 0.03 & 2.1  &  1.57  &   0.23 \\
NED01(old)&   0.0     &   0.0  &   100.0  &   0.0  &   0.0  &   100.0  & 0.02 & 2.2  &  1.56  &  -0.04 \\
\noalign{\smallskip}
\hline
\noalign{\smallskip}
NED02	  &	13.0  &   0.0  &   87     &   0.1  &   0.0  &   99.9   &  0.04  & 2.0     &  1.05  &  0.27 \\
NED02(old)&      0.0  &   0.0  &   100.0  &   0.0  &   0.0  &   100.0  &  0.03  &   2.2  &  1.12  &  -0.05 \\
\noalign{\smallskip}
\hline
\noalign{\smallskip}
\end{tabular}
\begin{minipage}[c]{18.0cm}
[1] Abundance by mass with Z$_{\sun}$=0.02 \\
\end{minipage}
\end{table*}

\section{Discussion}

In the \object{FM\,047-02} system, the NED01 galaxy is seen almost face-on and is clearly ringed, with two or three prominent smooth structures on the S and SE sides of the ring. No clearly prominent knots are visible in other regions of the ring, as detected in the Solitaire-candidate galaxies AM\,2152-592, AM\,2012-282, and AM\,1434-783. As can be seen in Fig.~\ref{fig_01}, the object shows two clear signatures of tidal perturbations: one main external ring and an internal loop-ring-like structure extending to the SW. 

NED01 resembles the collisional RE class of rings (empty rings) proposed by Theys \& Spiegel (\cite{ts76}). Recently, Mapelli \& Mayer (\cite{mm2012}) have investigated the formation of RE galaxies from N-body simulations. The RE galaxies are characterized by a ring that is empty in its interior, and apparently lacking the nucleus. The RE objects happen to have high star formation rates, suggesting that the density wave associated with the propagating ring triggers the formation of stars. From off-center collision simulations, Mapelli \& Mayer (\cite{mm2012}) demonstrate that the nucleus of the target galaxy is displaced from the dynamical center of the galaxy and placed within the ring. This result could explain the NED01 morphology. 

In both FM\,047-02 galaxies, along  the entire slit, the spectra show features that are characteristic of early-type objects. No star-forming regions and no nuclear ionization sources were detected. The two galaxies form a tidally bound system with a radial velocity difference of almost 240 km\,s$^{-1}$. 
The velocity dispersions $\sigma_{v}$ of 387 (NED01) and 285 km\,s$^{-1}$ (NED02) were estimated by cross-correlation with K and M star templates, with the statistical error in the range of $10\%-15\%$ ($XCSAO/RVSAO$, 
also tested with $XCOR/STSDAS$). We determined the dynamical masses for both galaxies based on the virial equation, assuming that both galaxies are at a distance of 322 Mpc. Using the Gemini pointing image and the Digitized Sky Survey one, we adopted the effective radii of 12\farcs7 and 11\farcs5 and velocity dispersions of 387 km\,s$^{-1}$ and 285 km\,s$^{-1}$, respectively. The calculated mass of NED01 (3.97$\times10^{11}$ M$_{\sun}$) is approximately twice that of the mass of NED02 (1.93$\times10^{11}$ M$_{\sun}$). Besides this, there are no prominent field galaxies within 9 arcmin of FM\,047-02, but there are several spherical objects that resemble dwarf galaxies within 2.5 arcmin and an elliptical-like object on the W of NED02 (with coordinates $\alpha$= 20$^{\rm h}$ 26$^{\rm m}$ 29\fs22 and $\delta$= -72\degr \,35\arcmin \,55\farcs0, 2000). In this context, the galaxy pair FM\,047-02 could be the main objects of a group of galaxies.

The observed radial-velocity distribution of NED01 shows some kinematically decoupled components, which are the products of the tidal interaction with NED02. These kinematic subsystems, displayed in Fig.~\ref{fig_02}, are unevenly distributed, including the very central region of this galaxy, of which center has a decoupled NE-section approaching us (on the side in the direction of NED02), and the almost rigid section including the very center between 0\farcs0 and 1\farcs25 approximately. The lower panel of Fig.~\ref{fig_02} displays the almost ``undisturbed" rotation curve in the interval $\pm$1\arcsec \,for NED01. This suggests that when the galaxy has a low degree of internal rotation, the tidal coupling should produce this U-type profile.

In this section we review the main kinematic structures for the \object{FM\,047-02} system based on the spectroscopy. When inspecting Figs.~\ref{fig_02} and ~\ref{fig_03} with respect to both heliocentric velocity centers, the following is suggested: (a) the core of NED02 is counter-rotating; (b) the external region of NED02 is kinematically decoupled from its central region (while  the NE section is approaching us, the SW section is receding from us); and (c) both sections of the NED01 ring are receding from us, along with the internal loop-ring. There are two phenomena inside the \mbox{U-shaped-base} of NED01: (1) the whole central region is displaced in the direction of the companion (NE direction), a phenomenon also seen in the aforementioned interacting pair HRG\,2304; (2) the kinematical center is also displaced in the NE direction.
 
The spectroscopic results corroborate the suggestion that both galaxies are early-type objects. It is important to emphasize that the fitting of stellar population pointed out a small, but non-negligible, fraction of young stars along the radius  of galaxy NED02 (about 10\%). On other hand, the galaxy NED01 is dominated by an old star population. No intermediate stellar populations have been identified (see Table \ref{synt_table}). Similar results have also been found for the peculiar pRG \object{HRG\,2304} and its companion NED01 (Wenderoth et al. \cite{wend11}), with a fraction of the young population along the radius of both galaxies (almost 9\%), but with some portion of intermediate stellar population.

Studies of the fundamental plane and studies based on absorption line spectroscopy favor a ``frosting" model in which early-type galaxies consist of an old base population with a small amount of younger stars (Trager et al. \cite{trager2000}; Gebhart et al. \cite{gebhardt2003}; and Schiavon \cite{schiavon2007}). In addition, data on early-type galaxies that exhibit strong ultraviolet excess, polycyclic aromatic hydrocarbon emission, and infrared excess are interpreted as a possible result of recent low-level star formation (Yi et al. \cite{yi2005}; Rich et al. \cite{rich2005}; Schawinski et al. \cite{scha2007}; Kaviraj et al. \cite{kaviraj2007}; Temi et al. \cite{temi2009}; Young et al. \cite{young2009} and Salim \& Rich \cite{sr2010}). Studies based on spatially resolved spectroscopy by Shapiro et al. (\cite{sha2010}) and Kuntschner et al. (\cite{kun2010}) have found that star formation in early-type galaxies happens exclusively in fast-rotating objects and occurs in two different contexts: (a) objects with a 
widespread young stellar population associated with a high molecular gas content; and (b) objects with disk and/or ring morphology. The latter could be explained by rejuvenation in previously quiescent stellar systems (Shapiro et al. \cite{sha2010}; Kuntschner et al. \cite{kun2010}), and seems to be suitable to explaining both galaxies of FM\,047-02.

The ring structure of NED01 was probably triggered by a wave of enhanced density moving outwards, resulting only in the redistribution of the old stellar population of this galaxy. This structure resembles the asymmetric distribution of the galaxy contents seen in the simulations from off-center collisions performed by Appleton \& Struck-Marcell (\cite{app_st1987}) and Mapelli \& Mayer (\cite{mm2012}).

\section{Conclusions}

In this work, we report observational results of the pRG NED01 and its companion NED02, both members of \object{FM\,047-02}. The galaxy NED01 was previously classified as a Solitaire-like galaxy by FAOA. Our work is based on spectroscopic observations in the optical. Below is a summary of our main results:

\begin{itemize}

\item The \object{FM\,047-02} system is composed of the ring galaxy NED01 and the elliptical-like galaxy NED02, (2MASX J20262950-7236443 and 2MASX J20263396-7236083, in NED, respectively). The projected separation between both galaxies' centers is almost 60 kpc. The velocity difference is close to $\Delta$v= 240 (km\,s$^{-1}$), confirming that both galaxies are interacting close companions.

\item Several spherical dwarf objects lie within 2.5 arcmin of \object{FM\,047-02}. In the NW direction of NED02,  an elliptical galaxy is close enough to be a field satellite, but there is no quoted redshift for this object in the literature. The galaxy pair FM\,047-02 seems to be surrounded by a group of dwarf objects.

\item The spectra of both galaxies resemble that of an early-type galaxy. No emission lines are detected in the observed sections of the NED01 ring, and no nuclear ionization sources were detected.

\item Both observed distributions of the rotation profile show some kinematically decoupled components. These are the results of the tidal interaction between both objects.  The discrete U-shaped rotation profile in NED01 ($\pm$5\arcsec) is thus direct observational evidence of tidal coupling, hence a direct observational signature of tidal friction in action with NED02.

\item Both galaxies are dominated by old stellar populations with ages between $2\times10^{9} <\rm t \leq 13\times10^{9}$ yr. About 13\% of young stars are estimated to contribute to the optical flux for NED02. 

\item The morphology, kinematical properties, the early-type characteristics  of the central region, the smoothness of the ring around NED01, and its off-center nucleus all characterize it as a true archetype of a Solitaire ring galaxy in an advanced stage of ring formation.

\end{itemize}

NED01 is a pRG where signatures of tidal perturbations are clearly visible. One ring and an internal ring-like structure are visible in Fig.~\ref{fig_01}. This remarkable system may have been generated by an off-center collision, as demostrated by N-body simulations by Mapelli \& Mayer (\cite{mm2012}), the disk of NED01 having given rise to the ring. A probably  early stage of this system could be represented by the galaxy pair \object{IC\,5364} \mbox{(\object{AM\,2353-291})}, a clearly interacting system with a velocity difference of \mbox{$\Delta$V=22 km\,s$^{-1}$}, and when in an advanced stage of dynamical evolution, by \mbox{\object{AM\,2145-543}}, which is probably the result of a off-center impact. We intend to conduct an exploratory study of \mbox{FM 047-02} by employing numerical N-body/hydrodynamical simulation to reconstruct its history and to predict the evolution of this tidal interaction.

\begin{acknowledgements}
      
This work was partially supported by the Ministerio da Ci\^{e}ncia, Tecnologia e Inova\c{c}\~{a}o (MCTI), Laborat\'{o}rio Nacional de Astrof\'{i}sica (LNA), and Universidade do Vale do Para\'{i}ba - UNIVAP. A. C. Krabbe  thanks the support of FAPESP, process 2010/01490-3. The paper is based on observations obtained at the Gemini Observatory, which is operated by the Association of Universities for Research in Astronomy, Inc., under a cooperative agreement with the NSF on behalf of the Gemini partnership: the National Science Foundation (United States), the Science and Technology Facilities Council (United Kingdom), the National Research Council (Canada), CONICYT (Chile), the Australian Research Council (Australia), Ministério da Ciência e Tecnologia (Brazil) and Ministerio de Ciencia, Tecnología e Innovación Productiva  (Argentina). The observations were performed under the identification number GS-2007A-DD-06. This publication makes use of data products from the Two Micron All Sky Survey, which is a joint project of 
the University of Massachusetts and the Infrared Processing and Analysis Center/California Institute of Technology, funded by the National Aeronautics and Space Administration and the National Science Foundation. This research made use of the NASA/ IPAC Infrared Science Archive, which is operated by the Jet Propulsion Laboratory, California Institute of Technology, under contract with the National Aeronautics and Space Administration.

\end{acknowledgements}

\end{document}